\documentclass[preprintnumbers,superscriptaddress,amsmath,amssymb,nofootinbib,twocolumn,showpacs]{revtex4-2}
\usepackage[utf8]{inputenc}
\usepackage{amsfonts}
\usepackage{graphicx}
\usepackage{geometry}
\usepackage{mathtools}
\usepackage{xcolor}

\usepackage[pdftex,
            breaklinks=true,%
            colorlinks=true,%
            linkcolor=red,
            urlcolor=blue,
            citecolor=blue,
            pdfauthor={Petac et al.},%
            pdftitle={Template for article in XXX}%
           ]{hyperref}

\newcommand{\M}[0]{M_\odot}

\newcommand{\ben}[0]{\begin{eqnarray}}
\newcommand{\een}[0]{\end{eqnarray}}
\newcommand{\nn}[0]{\nonumber}
\newcommand{\dd}[0]{{\rm d}}

\newcommand{\citefig}[1]{Fig.~\ref{#1}}
\newcommand{\citeeq}[1]{Eq.~\eqref{#1}}

\def\aap{A\&A}%
\def\apj{ApJ}%
\def\apjl{ApJ}%
\def\jcap{J. Cosmology Astropart. Phys.}%
\def\mnras{MNRAS}%
\def\nat{Nature}%
\def\physrep{Phys.~Rep.}%
\def\prd{Phys.~Rev.~D}%
\def\prl{Phys.~Rev.~Lett.}%

\begin{document}
\title{Microlensing constraints on clustered primordial black holes}
\author{Mihael Peta\v c}
\email{mihael.petac@ung.si}
\affiliation{University of Nova Gorica, Center for astrophysics and cosmology, Vipavska 11c, Ajdovščina, Slovenia}
\affiliation{Laboratoire Univers et Particules de Montpellier (LUPM), Universit\'e de Montpellier (UMR-5299) \& CNRS, Place Eug\`ene Bataillon, F-34095 Montpellier Cedex 05, France}
\author{Julien Lavalle}
\email{lavalle@in2p3.fr}
\author{Karsten Jedamzik}
\email{karsten.jedamzik@umontpellier.fr}
\affiliation{Laboratoire Univers et Particules de Montpellier (LUPM), Universit\'e de Montpellier (UMR-5299) \& CNRS, Place Eug\`ene Bataillon, F-34095 Montpellier Cedex 05, France}

\begin{abstract}
The discovery of black-hole-binary mergers through their gravitational wave (GW) emission has reopened the exciting possibility that dark matter is made, at least partly, of primordial black holes (PBHs). However, this scenario is challenged by many observational probes that set bounds on the relative PBH abundance across a broad range of viable PBH masses. Among these bounds, the ones coming from microlensing surveys lead to the strongest constraints in the mass range from $\sim 10^{-10}$ to a few M$_{\odot}$. The upper part of this range precisely corresponds to the mass window inside which the formation of PBHs should be boosted due to the QCD phase transition in the early Universe, which makes the microlensing probes particularly important. However, it has been argued that taking into account the inevitable clustering of PBH on small scales can significantly relax or entirely remove these bounds. While the impact of PBH clustering on the GW event rate has been studied in detail, its impact on the microlensing event rate has not yet been fully assessed. In this Letter, we address this issue, and show that clusters arising from isocurvature perturbations, that originating from PBHs formed from Gaussian initial curvature perturbations, do not alter the current microlensing constraints, as they are not sufficiently compact.
\end{abstract}
	
\maketitle
	
\preprint{LUPM:22-002}
	
\section{Introduction}
\label{sec:intro}
The advent of gravitational wave (GW) astronomy \cite{LigoVirgoCollabs2016} has opened a new window on the Universe. In particular, the peculiar mass range inferred from the first observations of binary-black-hole (BBH) mergers \cite{LigoVirgoCollabs2016,LigoVirgoCollabs2016a,LigoVirgoCollabs2019a} has been interpreted as a hint for a possible primordial-black-hole (PBH) origin of these events \cite{BirdEtAl2016,Clesse:2016vqa}. This revived the exciting idea that the still mysterious dark matter (DM), which pervades the Universe and is currently the pillar of the theory of structure formation \cite{Peebles1982,BardeenEtAl1986,SilkEtAl2012}, could be made, at least partly, of PBHs \cite{CarrEtAl1974,Chapline1975,BirdEtAl2016}.

However, this scenario has to confront several observational constraints that currently bound the PBH abundance along an extended range of masses (for recent reviews, see e.g.~\cite{CarrEtAl2020a,GreenEtAl2021a}). Among them, microlensing surveys provide a powerful probe in a very special mass range for PBHs, from a fraction of up to a few Solar masses \cite{Paczynski1986,Griest1991}. This mass range corresponds to PBHs that could have been formed during the QCD phase transition, when a drop of pressure decreases the overdensity threshold above which density fluctuations can collapse directly to black holes (BHs)~\cite{Jedamzik1997}. For approximately scale-invariant pre-existing density perturbations, one expects thus a peak in the mass function of PBHs around one Solar mass \cite{ByrnesEtAl2018}, which is within the reach of Galactic microlensing surveys \cite{MACHOEtAl2001,EROS2EtAl2007,OGLEEtAl2011}.
 

Current constraints limit PBHs to represent $f_{\rm pbh}\lesssim 0.1$ of the total DM abundance in this mass range. However, these limits have been derived assuming the PBHs are randomly distributed along a given DM density profile. Throughout the literature it has been suggested that the inevitable clustering of PBHs could significantly relax these bounds~\cite{ClesseEtAl2018a,CalcinoEtAl2018,CarrEtAl:2019}. In this Letter, we wish to explore in detail how such clustering affects the microlensing limits. We shall see that for Gaussian inflationary perturbations they are actually unlikely to change.



Where do these clusters come from? On large scales, the behavior of PBHs is indistinguishable from that of any viable cold dark matter (CDM) particle candidate~\cite{CarrEtAl2018}. However, on small scales there are two important differences compared to the particle hypothesis, i.e. the macroscopic PBH masses and their essentially vanishing initial peculiar velocities. For Gaussian perturbations  overdense regions exceeding the threshold for PBH formation are randomly distributed in space~\cite{Ali-Haimoud:2018dau,Desjacques:2018wuu}. This subsequently leads to Poisson fluctuations of PBH numbers in any given volume \cite{Meszaros1974,Meszaros1975}. These fluctuations represent the dominant source of isocurvature perturbations on small scales and inevitably seed the formation of PBH clusters already at very early times \cite{AfshordiEtAl2003}. However, due to small relaxation times, smaller clusters are subject to later evaporation.

The Poisson fluctuations in PBH numbers and their vanishing initial velocities lead to efficient formation of PBH binaries~\cite{AfshordiEtAl2003,RaidalEtAl2019,InmanEtAl2019}. These binaries are highly eccentric due the vanishing initial PBH peculiar velocities, and one expects a large BBH merger rate if PBHs make a significant fraction of CDM. In fact for $f_{\rm pbh}=1$, it  was thought that the merger rate of PBHs should be orders of magnitude larger than the one observed by LIGO/Virgo~\cite{Sasaki:2016jop}. However, these estimates did not take into account the presence of dense clusters, and PBH three-body interactions within, which significantly perturb the orbits and thus coalescence times of binaries. Detailed studies found that the merger rate could indeed be consistent with the observed one~\cite{Jedamzik2020,Jedamzik2021}, in case that most of the PBH binaries are actually residing in clusters. As of now, it seems that a $f_{pbh}=1$ with $M\sim M_{\odot}$ is marginally ruled out due to the coalescence of isolated and highly eccentric binaries~\cite{Jedamzik2021b}.


In this work we carefully examine the impact of PBH clustering on the interpretation of microlensing searches for massive compact halo objects (MACHOs). In our analysis, we pay particular attention to the differences in the probability distribution for the expected number of events, as well as the probability for recurring microlensing events, which were typically discarded from the observational samples. For concreteness, we provide quantitative results for the EROS-like microlensing survey~\cite{EROS2EtAl2007} of the Large Magellanic Cloud (LMC). However, the implications of our findings can be easily generalized to other similar Galactic surveys, such as those of MACHO~\cite{MACHOEtAl2000} and OGLE~\cite{OGLEEtAl2011}, or even Subaru/HSC observations of the Andromeda galaxy~\cite{NiikuraEtAl2019}.

\section{Microlensing searches for MACHOs}
\label{sec:microlensing}
Microlensing surveys search for temporary brightening of distant sources (e.g., stars) due to the passage of a MACHO near the line-of-sight (l.o.s.). In essence, this phenomenon is a particular case of strong lensing, where the angular separation between multiple images of the source is smaller than the resolution of the telescope (i.e. of order of microarcseconds, hence the name microlensing). The observed effect is the amplification of the source's brightness by a factor~\cite{Refsdal1964,Paczynski1986}:
\ben
\label{eqn:amplification}
A(u) = \frac{u^2 + 2}{u \sqrt{u^2 +4)}} \, ,
\een
where $u = R / R_{\rm E}$ is the displacement of the lens from the l.o.s. axis, expressed in units of Einstein radius,
\ben
R_{\rm E} = \sqrt{\frac{2r_{\rm S}(D_{\rm s}-D_{\rm l})D_{\rm l}}{D_{\rm s}}}\, , 
\een
where $r_{\rm S},D_{\rm s}$, and $D_{\rm l}$ are the Schwarzschild radius of the lens, the distance of the observer to the source, and the distance of the observer to the lens, respectively. The observed light curves are typically fitted using the following parameterization of $u$ \cite{RouletEtAl1997}:
\ben
\label{eqn:u_parametrization}
u(t) = \sqrt{u_0^2 + \frac{(t - t_0)^2}{t_{\rm E}^2}} \,
\een
where $u_0$ is the impact parameter of the lens, $t_0$ the time of maximum amplitude, while $t_{\rm E}$ is the characteristic crossing time, i.e. $t_{\rm E} = R_{\rm E} / v_\perp$ with $v_\perp$ being the lens transverse velocity with respect to the l.o.s.
	
The main quantity in predicting the observational signatures of MACHOs is the differential event rate $\dd \Gamma / \dd t_{\rm E}$. The latter quantifies the rate at which lenses enter the ``microlensing tube'', defined as the volume along the l.o.s. axis that leads to an amplification above a certain threshold $A_{\rm T} \equiv A(u_{\rm T})$ (for more details see, e.g.,~\cite{Griest1991} and Fig.~1 therein). $\dd \Gamma / \dd t_{\rm E}$ can be used to compute the expected number of microlensing events in a given survey with detection efficiency $\varepsilon(t_{\rm E})$ and total exposure $E$:
\ben
\label{eqn:N_expected}
\bar{N} = E  \int \dd t_{\rm E} \;\; \varepsilon(t_{\rm E}) \times \frac{{\rm d} \Gamma}{{\rm d} t_{\rm E}}(t_{\rm E}) \, .
\een

In microlensing studies, it is commonly assumed that PBHs follow a smooth and spherically symmetric Galactic halo density profile $\rho(r)$, where $r$ denotes the galactocentric distance, and have an isotropic Maxwellian velocity distribution with a dispersion $\sigma(r)$. In that case, the differential event rate can be computed by performing a single l.o.s. integral~\cite{Griest1991,MACHOEtAl1996}.

However, as argued in~\cite{ClesseEtAl2018a,CalcinoEtAl2018,Garcia-BellidoEtAl2018,CarrEtAl:2019}, these assumptions may no longer be justified if PBHs are bound into compact clusters, but rather should be applied to the clusters themselves and no longer to individual PBHs. This argument was used to evade microlensing constraints on PBH dark matter simply from the fact that (compact) cluster masses fall outside from the microlensing target mass range. It is straightforward and intuitively clear to write down the approximate condition for when a cluster should be treated as a single lens (compact cluster) and when as a compilation of spatially correlated individual lenses (extended cluster). Imagine a perfectly aligned observer-lens-source event where the bending of light rays by the Einstein angle $\Theta_{\rm E} = R_{\rm E}(M_{\rm cl})/D_{\rm l}$ is maximal, where $M_{\rm cl}$ is the cluster mass. These light rays pass approximately $\Theta_{\rm E}D_{\rm l} = R_{\rm E}(M_{\rm cl})$ from the center of the cluster and do not enter it if its radius is $r_{\rm cl} \ll R_{\rm E}(M_{\rm cl})$. Therefore, a cluster has to be treated as compact from the point of view of microlensing surveys when $r_{\rm cl} \ll R_{\rm E}(M_{\rm cl})$. We will see that for Gaussian perturbations this is not the case, as the clusters are orders of magnitude larger than the corresponding Einstein radius, i.e. $r_{\rm cl} \gg R_{\rm E}(M_{\rm cl})$. In such situations, characterized by low optical depth, accounting only for the PBH that is the closest to the l.o.s. is sufficient to model the net lensing effect~\cite{FleuryEtAl2020}.
\section{Some properties of PBH clusters}
\label{sec:clustering}	
For Gaussian primordial curvature perturbations the probability for the collapse into BHs is uniform in space and, therefore, results in initially randomly distributed PBHs. This unavoidably seeds Poissonian isocurvature perturbations on small scales, which later lead to the formation of PBH clusters. The overdensities associated with scales that on average contain $N$ PBHs are of the size of $\delta(N) \sim \Delta N / N = 1 / \sqrt{N}$. In the radiation dominated era, these fluctuations only grow logarithmically, however, after matter-radiation equality, they enter the linear growth regime and collapse. Throughout their subsequent evolution, smaller clusters evaporate, but are themselves aggregated into more massive structures. The typical evaporation time can be estimated as~\cite{BinneyEtAl2008}:
\ben
t_{\rm evap} \sim 140 \, t_{\rm relax} \sim \frac{14 \, N_{\rm pbh}}{\log N_{\rm pbh}} \; t_{\rm cross} \, ,
\een
where $t_{\rm relax}$ is the system relaxation time rewritten in terms of the number of PBHs within the cluster, $N_{\rm pbh}$, and the typical crossing time $t_{\rm cross} \sim r_{\rm cl} / v_{\rm cl}$ determined by the cluster size, $r_{\rm cl}$, and its virial velocity, $v_{\rm cl}$. At present time, one expects to find the largest abundance of PBH clusters with total masses $M_{\rm cl} \sim 10^4 \, \M$, consisting of $\sim 10^3-10^4$ individual PBHs with masses $\sim 1-10\,\M$, and with typical cluster sizes $r_{\rm cl} \sim 10 \, {\rm pc}$~\cite{Hutsi:2020sol,Jedamzik2020,Jedamzik2021}. Larger clusters, which are less dense, are made up of these basic building blocks, i.e. the of $M_{\rm cl} \sim 10^4 \, \M$ clusters, as these are the smallest structures that have not yet evaporated.
%
%
%
%
\section{Microlensing constraints for clustered PBH}
\label{sec:microlensing_constraints}
As mentioned above, if PBHs live in compact clusters, one can readily rescale the microlensing limits on the DM fraction in PBHs by considering single lenses of mass $M_{\rm cl}$. This can completely lift the current microlensing bounds on sub-Solar mass PBHs for $M_{\rm cl} \gg M_\odot$~\cite{ClesseEtAl2018a}. However, the condition for PBH clusters to act as individual lenses in surveys of the Magellanic Clouds is that $r_{\rm cl}  \ll {\rm 10\, au}  (M_{\rm cl}/\M)^{1/2}$ (for surveys targeting the Andromeda galaxy this ratio is larger, but only by a factor of $\sim 20$). This criterion is definitely {\em not} fulfilled for PBH clusters formed through the collapse of Gaussian primordial adiabatic perturbations~\cite{InmanEtAl2019}, whose typical features have been recalled above. Such dense PBH clusters can only form in the early Universe through some other mechanism (e.g.~\cite{BerezinEtAl1983,Hogan1984,BelotskyEtAl:2019}), or different initial conditions, such as non-Gaussian perturbations~\cite{YoungEtAl2020} (see also \cite{CarrEtAl2021b}).
We postpone a dedicated study of such scenarios to future work, however, we anticipate that substantial fine-tuning of the corresponding PBH production mechanism would be required to significantly relax the microlensing bounds, while still obeying other observational constraints. Therefore, in the following, we focus on the effect of extended PBH clusters on the microlensing limits derived from the surveys of the LMC.

\subsection{Probability distribution for the number of observed events}
\label{ssec:pdf}
To carry out the statistical analysis of the event rate, we resort to dedicated Monte-Carlo (MC) simulations of microlensing signals produced by clustered PBHs. While semi-analytical estimates are possible for sparsely distributed sources, 
the crowded stellar fields of nearby galaxies make it necessary to perform a numerical convolution of associated complex stellar surface densities with the probability distribution of PBH cluster positions, rendering it analytically intractable.
	
\subsubsection{Monte-Carlo simulations}
\label{sssec:mc}
Our MC simulations consist in drawing $N_{\rm cl} \sim {\rm Pois}(\lambda_{\rm LMC})$ PBH clusters, whose positions and velocities are distributed according to a chosen global model of the Galactic DM halo. The latter also defines $\lambda_{\rm LMC} = M_{\rm l.o.s.} / M_{\rm cl}$, i.e. the expected number of PBH clusters within the observed volume ($M_{\rm l.o.s.}$ corresponds to the DM mass contained in a cone with the apex at the position of the observer and the base in the approximate plane of the stellar source population, encompassing the observed area in the sky). We additionally assume that all PBH clusters have the same mass $M_{\rm cl}$, as justified above. 
For each of the clusters, we draw a number of microlensing events from a Poisson distribution with the appropriate mean. The latter is determined by computing the differential event rate, $\dd \Gamma$, from its standard expression (see, e.g., Eq.~8 in~\cite{Griest1991}). We further assume that all PBHs have equal masses, $M_{\rm PBH} = 1 \, M_\odot$, and that the velocity dispersion of PBHs inside a cluster is negligible with respect to the cluster velocity in the Galactic halo, such that the velocity distribution of the cluster members can be modeled as a delta function picked at the cluster's peculiar velocity $\vec{v}_{\rm p}$. Thus, the differential event rate $\dd \Gamma$ (integrated over angles) can be written as:
\ben
\label{eqn:differental_event_rate}
\dd \Gamma = 2\,n_{\rm PBH}(x) \, u_{\rm T} \, v_{\perp} \, R_{\rm E}(x) \, \dd x \; ,
\een
where $n_{\rm PBH}(x)$ is the PBH number density at position $x$ along the l.o.s. Consequently, the Einstein crossing time of each event associated with this cluster is fixed to $t_{\rm E} = R_{\rm E}(x) u_{\rm T} / v_{\perp}$. The corresponding differential event rate {\em inside a cluster} for a l.o.s with impact parameter $b$ can then be expressed as the l.o.s. integral:
\ben
&\frac{{\rm d}\Gamma}{{\rm d}t_{\rm E}}(b, t_{\rm E}) =  \frac{2 v_\perp u_{\rm T} R_{\rm E}(x_{\rm cl})}{M_{\rm PBH}}\, \delta\left(\frac{R_{\rm E}(x_{\rm cl}) u_{\rm T}}{v_\perp} - t_{\rm E}\right) \nonumber \\ 
& \times \int_{x_{\rm cl} - \Delta x}^{x_{\rm cl} + \Delta x} {\rm d}x \; \rho_{\rm cl}\left(\sqrt{(x-x_{\rm cl})^2 + b^2}\right) \, ,
\een
where $x_{\rm cl}$ denotes the position of the PBH cluster along the l.o.s., $\Delta x\equiv\sqrt{r_{\rm cl}^2-b^2}$ the half-length of the l.o.s. segment penetrating the cluster, and $\rho_{\rm cl}(r)$ the radial density profile of the cluster (we assume that the PBH clusters are spherically symmetric). The expected number of microlensing events for the entire cluster is obtained by convoluting the above expression with the surface density of stellar targets, $\Sigma_\star(\textbf{q})$, over the area of the cluster (neglecting the small angle correction as $r_{\rm cl} \ll  x_{\rm cl}$):
\ben
\frac{{\rm d}\Gamma_{\rm cl}}{{\rm d}t_{\rm E}}(t_{\rm E}) & = & \int_{0}^{r_{\rm cl}} {\rm d}b \,\, b \,\frac{{\rm d}\Gamma}{{\rm d}t_{\rm E}}(b, t_{\rm E}) \nonumber \\ 
& \times & \int_0^{2\pi} {\rm d} \varphi \,\, \Sigma_\star\left(\textbf{q}_{\rm cl} - \textbf{q}(b, \varphi)\right) \; ,
\een
where $\textbf{q}_{\rm cl}$ denotes the position of the cluster on the celestial sphere and $\textbf{q}(b, \varphi)$ the displacement corresponding to given values of $b$ and $\varphi$. This differential rate translates into the expected number of lensing events induced by the $i^{\rm th}$ cluster:
\ben
\bar{N}_{{\rm e},i} = t_{\rm obs} \, \int {\rm d}t_{\rm E} \,\, \varepsilon(t_{\rm E}) \times \frac{d\Gamma_{{\rm cl},i}}{dt_{\rm E}}(t_{\rm E}) \;\; .
\een
Finally, we draw the number of events due to each cluster, $N_{{\rm e},i} \sim {\rm Pois}(\bar{N}_{{\rm e},i})$, and sum them up to obtain the total number of events in a given simulation run:
\ben
N_{\rm obs} = \sum_{i=1}^{N_{\rm cl}} N_{{\rm e},i} \; .
\een
Note that the stochastic distribution of PBH clusters in the observational cone and the uneven distribution of target stars on the celestial sphere lead to $N_{\rm obs}$ that does not trivially follow a Poisson distribution. This explains the need for performing dedicated MC simulations.
\subsubsection{Results}
\label{sssec:res}
For concreteness, here we apply our method to study the distribution of microlensing events in the EROS-2 survey of the LMC~\cite{EROS2EtAl2007}. In particular, we adopt the corresponding target star distribution and detection efficiency, while for the Milky Way DM halo we assume the same cored DM halo model -- for details see~\cite{EROS2EtAl2007}. 
	
The MC simulations allow us to determining the probability distribution for the total number of observed events, $P_{\rm obs}(N_{\rm obs})$, which we report in \citefig{fig:prob_dist} for various $M_{\rm cl}$ while keeping $r_{\rm cl} = 10 \, {\rm pc}$~\footnote{In the scenario where PBHs are formed from Gaussian curvature perturbations, the adopted value of $r_{\rm cl}$ corresponds to clusters with $M_{\rm cl} \sim 10^4 \, M_\odot$, while more massive clusters are expected to be larger and less dense, since $r_{\rm cl} \propto M_{\rm cl}^{5/6}$. However, for $M_{\rm cl} \gtrsim 10^7 \, M_\odot$ this would imply extremely large clusters with high chance of spatial overlap, and, therefore, we conservatively keep $r_{\rm cl} = 10 \, {\rm pc}$ for all the considered values of $M_{\rm cl}$. Note that despite this choice even clusters with $M_{\rm cl} = 10^8 \, M_\odot$ still acts as extended clusters.}. First thing to note is that the mean number of events, $\bar{N}_{\rm obs} \approx 23$, is always the same, regardless of the cluster mass, as it only depends on the total number of PBHs within the probed volume, i.e., $M_{\rm l.o.s.} / M_{\rm PBH}$. From the same figure it can also be seen that $P_{\rm obs}(N_{\rm obs})$ is virtually indistinguishable from the standard result obtained for unclustered PBH if $M_{\rm cl} \lesssim 10^6 \, M_\odot$. This comes from the fact that for a large number of individual clusters, fluctuations about individual l.o.s.'s average out and $P_{\rm obs}(N_{\rm obs})$ simply reduces to Poisson distribution (for $M_{\rm cl} = 10^6 \, M_\odot$ one expects $N_{\rm cl} \sim 1300$ within the field of view). However, for higher cluster masses,  $P_{\rm obs}(N_{\rm obs})$ becomes increasingly skewed toward low $N_{\rm obs}$, but with a thin tail reaching much larger values. For the highest PBH cluster mass considered in this work, $M_{\rm cl} = 10^8 \, M_\odot$, $P_{\rm obs}(N_{\rm obs})$ peaks at $N_{\rm obs} = 8$, but has an extremely long tail, stretching to hundreds of events. This would lead to very different exclusion limits on the allowed fraction of DM in the form of PBHs. For example, one could not exclude (at 95\% confidence level) DM being fully made of PBH clusters of masses $M_{\rm cl} = 10^8 \, M_\odot$ if more than 3 microlensing events were detected, while for smoothly distributed PBH the threshold value shifts to 15. This clearly demonstrates the importance of taking into the account correlations along neighboring l.o.s.'s in the case of PBH clusters with $M_{\rm cl}$ approaching $M_{\rm l.o.s.}$.

While the presented results were obtained for an EROS-like survey of the LMC, we wish to emphasize that analogous conclusions are applicable also to other microlensing surveys aimed toward the Magellanic Clouds, as well as the Andromeda galaxy. Indeed, extended PBH clusters can significantly affect $P_{\rm obs}(N_{\rm obs})$ only if their individual masses represent a large fraction of the expected DM mass within the observational cone. Otherwise, the fluctuations in the expected number of the events along individual l.o.s.'s of each realization average out and the predicted signal becomes very similar to that of smoothly distributed PBH.
	
\begin{figure}
\includegraphics[width=\linewidth]{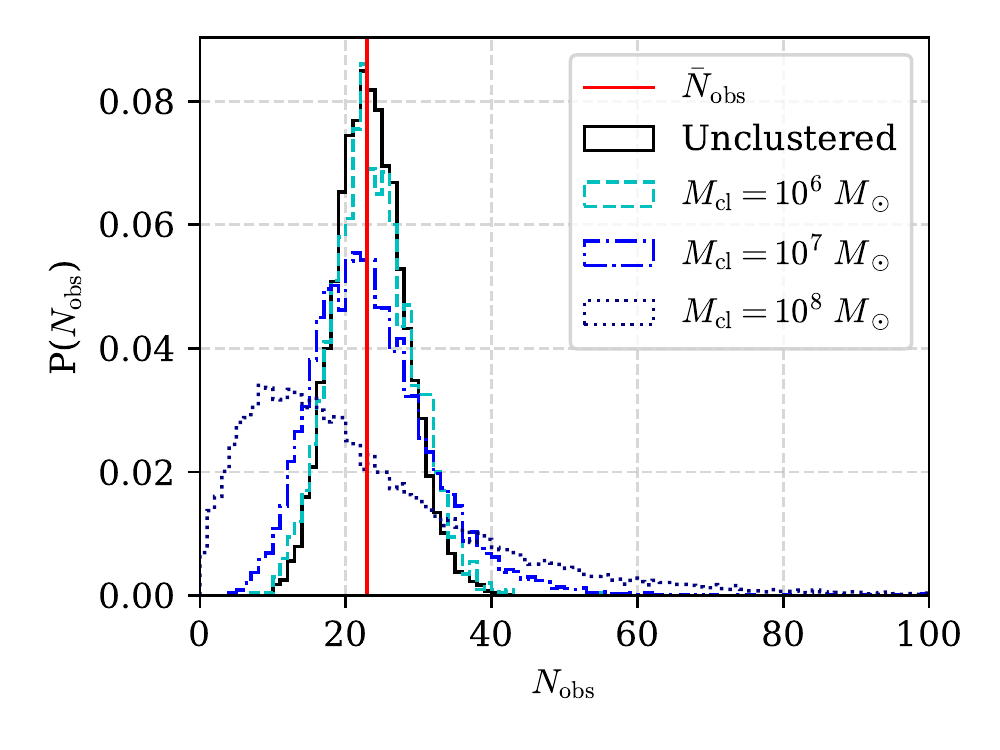}
\caption{Probability distribution functions for the number of observed microlensing events $N_{\rm obs}$ in EROS-2-like survey of the LMC. The black curve shows the standard result assuming smoothly distributed PBHs, while color curves represent our results for PBHs bound in clusters of different masses. The vertical red line shows the mean number of events, $\bar{N}_{\rm obs}$, which is independent of PBH clustering.}
\label{fig:prob_dist}
\end{figure}

\subsection{Recurrence of events}
\label{sec:formalism_recurrence}
Microlensing events are extremely rare due to the smallness of the volume that has to be crossed by a compact object to produce an observable amplification. Therefore, multiple microlensing events from the same source are conventionally attributed to its intrinsic variability rather than microlensing. This view has been challenged in several works proposing that highly clustered PBHs could lead to recurring amplifications along the same l.o.s.~\cite{ClesseEtAl2018a,CalcinoEtAl2018}. In the following, we critically reexamine this hypothesis, first focusing on the effect of small separation distances between individual PBHs in extremely compact clusters and, subsequently, on the possibility of recurring events due to PBH binaries, which are expected to be efficiently formed in the early Universe.

\subsubsection{Small PBH separation distances}
\label{sec:formalism_recurrence_dense}
For dense PBH clusters, the dynamical time is typically orders of magnitude smaller than their ages. Therefore, they should be virialized and the total energy of the cluster can be expressed as $E_{\rm cl} = -N M_{\rm pbh} v_{\rm cl}^2/2 = - G (N)^2/(4 r_{\rm cl})$~\cite{BinneyEtAl2008}, where $v_{\rm cl}$ is root-mean-square velocity of an individual PBH of mass $M_{\rm pbh}$, and $N$ is the number of PBHs within the cluster. This gives $v_{\rm cl}^2=GNM_{\rm pbh}/(2\,r_{\rm cl})$. On the other hand, the separation distance between the PBH can be estimated as $\Delta r \approx (V_{\rm cl}/N)^{1/3} = r_{\rm cl} (4 \pi/(3 N))^{1/3}$. From this, the typical time required for a PBH to cross $\Delta r$ reads:
\ben
\Delta t \approx \frac{\Delta r}{v_{\rm cl}} = \left(\frac{4 \pi}{3 N}\right)^{1/3} \left( \frac{2 r_{\rm cl}^3}{G N M_{\rm pbh}}\right)^{1/2} \;
 .
\label{Deltat} 
\een
By adopting a conservative approximation, $r_{\rm cl} \geq \sqrt{N} R_{\rm E}(M_{\rm pbh})$, for which the extended-cluster condition is only marginally fulfilled, we get:
\ben
\Delta t \gtrsim 9.8 \; {\rm yr} &  \left(\frac{x(1-x)}{0.5}\right)^{3/4} \left(\frac{D_{\rm s}}{55 \; {\rm kpc}}\right)^{3/4} \nn \\
& \times \left(\frac{M_{\rm pbh}}{M_\odot}\right)^{1/4} \times \left(\frac{10^4}{N}\right)^{1/12} \; ,
\een
where\footnote{Without impacting our conclusions below, for less dense clusters $v_{\rm p}$ should be taken in \citeeq{Deltat}.} $x = D_{\rm l}/D_{\rm s}$. While this is comparable to the duration of a typical microlensing survey, it is important to note that most of the events would not produce light curves consistent with \citeeq{eqn:amplification} if $r_{\rm cl} \sim \sqrt{N} R_{\rm E}(M_{\rm pbh})$, as the light rays would likely be affected by multiple PBHs. Therefore, in reality, $\Delta t$ has to be significantly (orders of magnitude) larger than the above value, making recurring events due to small PBH separation distances very unlikely, even for clusters consisting of an overwhelming number of sub-Solar mass PBHs.
\subsubsection{Binary PBHs}
\label{sec:formalism_recurrence_binary}
PBH binaries are efficiently formed in the early Universe around matter-radiation equality. A large fraction of these binaries survive three-body interactions in clusters \cite{RaidalEtAl2019,InmanEtAl2019,Jedamzik2020} and could potentially lead to recurring events \cite{ClesseEtAl2018a,CalcinoEtAl2018,Garcia-BellidoEtAl2018}. We can derive a lower bound on the time separation of microlensing events under the assumption that one of such pairs happens to lie along a given l.o.s. aligned with its orbital plane.
The latter can be simply estimated as one half of the orbital period of a two-body system:
\ben
\Delta t = \frac{T}{2} = \pi \sqrt{\frac{a^3}{G \mu}} \; ,
\een
where $a$ is the orbital semi-major axis and $\mu$ is the reduced mass of the PBH. To minimize $\Delta t$, we assume that both PBHs have the same mass, $\mu = M_{\rm pbh} / 2$, while we set $a \geq 2 R_{\rm E}(M_{\rm pbh})$, so that the single-lens assumption in \citeeq{eqn:amplification} is still marginally applicable. This yields:
\ben
\Delta t \geq 116 \; {\rm yr} & \left(\frac{x(1-x)}{0.5}\right)^{3/4} \left(\frac{D_{\rm s}}{55 \; {\rm kpc}}\right)^{3/4} \nn \\
& \times \left(\frac{M_{\rm pbh}}{M_\odot}\right)^{1/4} \; .
\een
This is again very conservative, since only a fraction of PBHs should be in the form of binaries with the orbital plane aligned along the l.o.s. and, additionally, one should have $a \gg 2 R_E$ to observe light curves that are not distorted by the presence of the second object (such anomalies have already been detected in a binary BH lens~\cite{OGLEEtAl2007}). Therefore, we conclude that PBH binaries are also very unlikely sources of recurring events, even for $M_{\rm pbh} \ll M_\odot$.
\section{Summary and conclusions}
\label{sec:concl}
In this work we have carefully examined the impact of PBH clustering on the interpretation of microlensing searches for MACHOs. Heuristically, it has often been argued that microlensing constraints on DM in the form of PBHs can be significantly relaxed, or even entirely removed, if one accounts for the fact that PBHs are bound into clusters. However, our dedicated MC simulations of EROS-2-like experiment show that microlensing limits can be lifted only if clusters are extremely compact,
$r_{\rm cl}\ll R_E(M_{\rm cl})$, 
or very massive, $M_{\rm cl} \gtrsim 10^8 \, M_\odot$. In the most commonly considered scenarios, where PBHs are formed through the collapse of Gaussian primordial curvature perturbations, both of these conditions are far from being fulfilled. Similar conclusions also hold for other microlensing surveys of the Magellanic Clouds, as well as of the Andromeda galaxy.
While in models that predict PBH formation from other initial conditions or through topological defects it may be possible to produce sufficiently dense PBH clusters to avoid the microlensing limits, further work is needed to establish their survival to the present days.

Similarly, it has been argued that PBH clustering can lead to recurring microlensing events which have been spuriously discarded from the observational samples. However, using simple analytical estimates, we have shown that recurring microlensing events due to dense PBH clusters or PBH binaries should occur on significantly longer timescales than the typical duration of microlensing surveys. Therefore, it is unlikely that any of the observed recurring events were in fact of a PBH origin.

\begin{acknowledgments}
We would like to thank Sébastien Clesse for interesting discussions during the preparation of this work, as well as Anne Green and Matthew Gorton for valuable feedback regarding the first e-print version of this work. We acknowledge support from the ANR project ANR-18-CE31-0006 ({\em GaDaMa}), the national CNRS-INSU programs PNHE and PNCG, and European Union's Horizon 2020 research and innovation program under the Marie Sk\l{}odowska-Curie grant agreement N$^{\rm o}$ 860881-HIDDeN. MP acknowledges partial support MP acknowledge partial support from the ARRS grant no. P1-0031. The authors are grateful for the computational resources and technical support provided by the IT department of LUPM.
\end{acknowledgments}

\bibliography{pbh_clusters} 

\end{document}